\documentclass[aps,pra,reprint]{revtex4-1}
\usepackage{graphicx}
\usepackage{amsmath}
\usepackage{amsfonts}
\usepackage{braket}
\usepackage{array}
\usepackage{bm}
\usepackage{bbm}
\newcommand{\bd}{b^\dagger}
\newcommand{\brd}{\bar{b}^\dagger}
\newcommand{\vrx}{\bar{v}}
\newcommand{\urc}{\bar{u}^*}
\newcommand{\vrc}{\bar{v}^*}
\newcommand{\vaca}{\Ket{0_a}}
\newcommand{\vacb}{\Ket{0_b}}

\newcommand{\brc}{\bar{\beta}^*}
\newcommand{\minus}{\scalebox{0.6}[1.0]{$-$}}
\renewcommand{\Re}{\mathrm{Re}}

\begin{document}
		\title{Spectrum and normal modes of non-hermitian quadratic boson operators} \author{Javier Garcia$^{1}$ and R.\ Rossignoli$^{1,2}$}
		\affiliation{$^1$IFLP-CONICET and Depto.\ de F\'{\i}sica, Universidad Nacional de La Plata,\\ C.C.67, La Plata 1900, 
			Argentina\\$^2$Comisi\'on de Investigaciones Cient\'{\i}ficas (CIC), La Plata (1900), Argentina}
\begin{abstract}
	We analyze the spectrum and normal mode representation  of general quadratic bosonic forms $H$ not necessarily hermitian. 
	It is shown that in the one-dimensional case such forms exhibit  either an harmonic regime where both $H$ and $H^\dagger$ 
	have a discrete spectrum with biorthogonal eigenstates, and  a coherent-like regime where either $H$ or $H^\dagger$ have 
	a continuous complex two-fold degenerate spectrum, while its adjoint has no convergent eigenstates. 
	These regimes reflect the nature of the pertinent normal 
	boson operators. Non-diagonalizable cases as well critical boundary sectors separating these regimes are also analyzed. 
	The extension to $N$-dimensional quadratic systems is as well discussed. 
\end{abstract}
\maketitle
\section{Introduction}
The introduction of parity-time ($\mathcal{PT}$)-symmetric Quantum Mechanics 
\cite{bender:1998:prl,bender:2007:rpp} has  significantly enhanced the interest  in non-hermitian Hamiltonians. 
When possessing $\mathcal{PT}$ symmetry, such  Hamiltonians can still exhibit a real spectrum if the symmetry is 
 unbroken in all eigenstates,  undergoing a transition to a regime with complex eigenvalues when the symmetry becomes 
 broken \cite{bender:1998:prl,bender:2007:rpp}. A generalization based on the concept of pseudohermiticity was then developed  \cite{mostafazadeh:2002:jmp_0,*mostafazadeh:2002:jmp_b,*mostafazadeh:2002:jmp_c,
	mostafazadeh:2010:ijgmmp,bagarello:book},  which provides a complete characterization of diagonalizable Hamiltonians 
with real discrete spectrum and is equivalent to the presence of an antilinear  symmetry. A similar approach had been already 
put forward  in \cite{scholtz:1992:aop} in connection with the non-hermitian bosonization of angular-momentum and fermion operators 
introduced by Dyson  \cite{dyson:1956:pr,*dyson:1956b:pr,janssen:1971:npa}. An   equivalent formulation of the general formalism 
based on biorthogonal states can also be made \cite{mostafazadeh:2010:ijgmmp,brody:2013:jpa,brody:2016:jpa}. 

Non-hermitian Hamiltonians were first introduced as  effective Hamiltonians for describing open quantum systems \cite{Feshbach:1958:aop}. 
Non-hermitian Hamiltonians with ${\cal PT}$ symmetry have recently provided successful effective descriptions of diverse systems and processes, 
specially in open regimes with balanced gain and loss. Examples are laser absorbers \cite{longhi:2010:pra}, 
ultralow threshold phonon lasers \cite{jing:2014:prl}, defect states and special beam dynamics in optical lattices 
\cite{regensburger:2013:prl,*makris:2008:prl}  and other related optical systems 
\cite{guo:2009:prl,rueter:2010:np,*lin:2011:prl,*feng:2011:science,*schnabel:2017:pra}. 
$\mathcal{PT}$-symmetric properties have been also observed and investigated in simulations of quantum circuits based on 
nuclear magnetic resonance \cite{zheng:2013:ptrsa}, superconductivity experiments \cite{rubinstein:2007:prl,chtchelkatchev:2012:prl}, 
microwave cavities \cite{bittner:2012:prl}, Bose-Einstein condensates \cite{kreibich:2016:pra,*schwarz:2017:pra},  
spin systems \cite{zhang:2013:pra,*zhang:2017:pra,*li:2016:pra}, and vacuum fluctuations \cite{pendharker:2017:pra}. 
Evolution under time-dependent non-hermitian Hamiltonians has also been discussed in  
\cite{znojil:2008:prd,*znojil:2009:sigma,znojil:2017:aop,*fring:2017:pra}.

Of particular interest are non-hermitian Hamiltonians which are quadratic in coordinates and momenta, or equivalently, 
boson creation and annihilation operators. They include the so-called 
Swanson models \cite{swanson:2004:jmp,jones:2005:jpa}, based on one-dimensional $\mathcal{PT}$-symmetric Hamiltonians with real spectra, 
which have been examined and extended in different ways  
\cite{scholtz:2006:jpa,*musumbu:2007:pra,sinha:2007:jpa,*sinha:2009:jpa,bagarello:2010:pla,*fring:2015:jmp,*fring:2017:ijmp,fring:2016:pra}. 
Effective quadratic non-hermitian Hamiltonians have also arisen in the description of LRC circuits with balanced gain and loss 
\cite{ramezani:2012:pra,*fernandez:2016:aop_0, *schindler:2011:pra}, coupled optical resonators \cite{bender:2013:pra,bender:2013:pra},  
optical trimers \cite{xue:2017:oe} and the interpretation of the electromagnetic self-force \cite{bender:2015:jmp}. 

The aim of this article is to examine the normal modes, spectrum and eigenstates of general, not necessarily hermitian, 
quadratic bosonic forms  in greater detail, 
extending the methodology of \cite{rossignoli:2005:pra_0, rossignoli:2009:pra_0} to the present general situation.  
Such quadratic forms can represent basic systems like a harmonic oscillator with a discrete spectrum, a free particle Hamiltonian 
with a continuous real spectrum, the square of an annihilation operator, in which case it has a continuous complex spectrum with 
coherent states \cite{glauber:1963:pr} as eigenvectors, 
and the square of a creation  operator, in which case it has no convergent eigenstates. 
We will here show that a general quadratic one-dimensional form belongs  essentially to one of these previous categories,
 as determined by the nature of the normal boson operators, i.e., as whether one, both or none of them 
 possesses a convergent vacuum. 
  Explicit expressions for eigenstates 
 are provided, together with an analysis of border and ``nondiagonalizable'' regimes. 
The extension to $N$-dimensional quadratic systems is then also  discussed.   

\section{The one-dimensional case}
\subsection{Normal mode representation}
We consider a  general quadratic  form in standard boson creation and annihilation operators $a, a^\dagger$ ($[a,a^\dagger]=1$), 
\begin{eqnarray}
	H &=& A \left( a^\dagger a + \frac{1}{2} \right) + \frac{1}{2}\left(B_+ a^{\dagger\,2} + B_- a^2 \right)\label{H1}\\
&=&\frac{1}{2}\begin{pmatrix} a^\dagger&a\end{pmatrix}\mathcal{H}\begin{pmatrix}a\\a^\dagger\end{pmatrix}\,,
\quad \mathcal{H} = \begin{pmatrix} A & B_+ \\ B_- & A \end{pmatrix}\,,
	\label{eqn:H}
\end{eqnarray}  
where $A$ and $B_{\pm}$ are in principle arbitrary complex numbers. By extracting a global phase  we can  always assume, nonetheless,  
$A$ real non-negative ($A\geq 0$),  while by  a phase transformation $a\rightarrow e^{i\phi}a$, 
$a^\dagger\rightarrow  e^{-i\phi} a^\dagger$, we can set equal phases on $B_{\pm}$, such that $B_\pm=|B_\pm|e^{i\theta}$.    
The hermitian case corresponds to ${\cal H}$ hermitian and the original Swanson Hamiltonian to $B_{\pm}$ real \cite{swanson:2004:jmp}. 

Our first aim is to write $H$ in the normal form 
\begin{equation}
H = \lambda \left( \brd b + \frac{1}{2}\right)\,,
\label{eqn:H_diag}
\end{equation}
where $b$, $\bar{b}^\dagger$ are related to $a$ and $a^\dagger$ through a generalized Bogoliubov transformation 
\begin{equation}
		b = u a + v a^\dagger,\quad
		\brd = \vrc a + \urc a^\dagger\,.
	\label{eqn:bogoliubov}
\end{equation}
Here $\brd$ may differ from $b^\dagger$ although they still  satisfy the bosonic commutation relation 
\begin{equation}[b,\brd] = 1\,,\end{equation} which implies   
\begin{equation}u\urc - v\vrc = 1\,.\label{det1}\end{equation}
If ${\cal H}$ is hermitian and positive definite ($|B_{\pm}|<A$), such that $H$ represents a stable 
bosonic mode, we can always choose $u,v,\bar{u}$ and $\bar{v}$   
such that $\bar{b}^\dagger=b^\dagger$. This choice  is no longer feasible in the general case. 

The transformation  (\ref{eqn:bogoliubov}) can be written as 
\begin{equation} 
\begin{pmatrix}b\\{\brd}\end{pmatrix}={\cal W}\begin{pmatrix}a\\a^\dagger\end{pmatrix}\,,\;\;
{\cal W}=\begin{pmatrix}u&v\\\vrc&\urc\end{pmatrix}\,,
\end{equation}
with  ${\cal W}$ satisfying ${\rm Det}\,{\cal W}=1$. We can then rewrite $H$ as 
\begin{eqnarray}
H&=&\frac{1}{2}\begin{pmatrix} \brd&b\end{pmatrix}\mathcal{H}'\begin{pmatrix}b\\\brd\end{pmatrix}\,,\label{Hb0}\\
\mathcal{H}' &=& {\cal M}{\cal W}{\cal M}{\cal H}{\cal W}^{-1}=\begin{pmatrix}A'&B'_+\\B'_-&A'\end{pmatrix}\,, \label{Hb}\end{eqnarray}
where $A'=A(u\urc+v\vrc)-B_+u\vrc-B_-\urc v$, $B'_+=B_+u^2+B_-v^2-2Auv$,   $B'_-=B_-{\urc}\,^2+B_+\vrc\,^2-2A\urc\vrc$ and 
\begin{equation}
{\cal M}=
\begin{pmatrix}1&0\\0&-1\end{pmatrix}\,.
\label{M}\end{equation}

It is then seen from Eq.\ (\ref{Hb}) that a {\it diagonal} $\mathcal{H}'$ ($B'_{\pm}=0$, $A=\lambda$) 
and hence  a diagonal representation   (\ref{eqn:bogoliubov}) 
can  be obtained {\it if and only if\,} i) the matrix 
\begin{equation}{\cal M}{\cal H}=\begin{pmatrix}A&B_+\\-B_-&-A\end{pmatrix},\label{MH}\end{equation}
whose eigenvalues are $\pm \lambda$ with 
\begin{equation}
\lambda=\sqrt{A^2-B_+B_-}\,,\label{la}
\end{equation}
is {\it diagonalizable},  i.e.\ $\lambda\neq 0$ if  ${\rm rank}({\cal H})>0$, and ii) ${\cal W}^{-1}$  
is a  matrix  with unit determinant diagonalizing ${\cal M} {\cal H}$,  
such that ${\cal W}{\cal M}{\cal H} {\cal W}^{-1}=\lambda {\cal M}$ and  ${\cal H}'=\lambda \mathbbm{1}$. 
 For instance, assuming $\lambda\neq 0$, we can set 
\begin{equation}
\begin{aligned}
&u=\urc=\sqrt{\frac{A+\lambda}{2\lambda}},\;\;\\
v=\sqrt{\frac{A-\lambda}{2\lambda}}&\sqrt{\frac{B_+}{B_-}}\,,\;\;\bar{v}^*=
\sqrt{\frac{A-\lambda}{2\lambda}}\sqrt{\frac{B_-}{B_+}}\,,\;\;
\end{aligned}
\label{eqn:parametros}
\end{equation}
where signs of $v$, $\vrc$ are such that $2\lambda u\vrc=B_-$, $2\lambda \urc v=B_+$. Any  further rescaling 
$b\rightarrow \alpha b$, $\brd\rightarrow \alpha^{-1}\brd$, $\alpha\neq 0$, remains feasible, since  
it will not affect their commutator nor Eq.\  (\ref{eqn:H_diag}), although the choice (\ref{eqn:parametros}) 
directly  leads to $\bar{b}^\dagger=b^\dagger$ when ${\cal H}$ is hermitian and positive definite 
(in which case $0<\lambda\leq A$).   
Eqs.\ (\ref{eqn:parametros}) remain also valid for $B_+\rightarrow 0$ or $B_-\rightarrow 0$, in which case 
$\lambda\rightarrow A$,  $u=\urc\rightarrow 1$ and  $(v,\vrc)\rightarrow (0,\frac{B_-}{2A})$ or $(\frac{B_+}{2A},0)$. 

If no further conditions are imposed on $b, \bar{b}^\dagger$, the sign chosen for $\lambda$ 
is irrelevant, since (\ref{eqn:H_diag}) can be rewritten as  $-\lambda(\bar{b}^{'\dagger}b'+\frac{1}{2})$ 
for $\bar{b}^{'\dagger}=-b$,  $b'=\brd$ (also satisfying $[b',\bar{b}^{'\dagger}]=1$). The sign can be fixed 
by imposing the condition that $b$ (rather than $\brd$) has a proper vacuum, as discussed in the next section, 
in which case the right choice for $A\geq 0$ is ${\rm Re}(\lambda)\geq 0$. 

The matrix ${\cal M} {\cal H}$ determines the commutators of $H$ with $a$ and $a^\dagger$, $[H,a]=-Aa-B_+a^\dagger$, 
 $[H,a^\dagger]=Aa^\dagger+B_-a$:  
\begin{equation} 
	[H,\begin{pmatrix}a\\a^\dagger\end{pmatrix}]=-{\cal M}{\cal
	H}\begin{pmatrix}a\\a^\dagger\end{pmatrix}\,.
	\label{eqn:conmuta}
\end{equation}
The normal boson operators $b,\bar{b}^\dagger$ satisfying 
(\ref{eqn:H_diag}) are then those  diagonalizing this semialgebra: 
\begin{equation}[H,b]=-\lambda b,\;\;[H,\brd]=\lambda \brd\,.\label{cb}\end{equation}
Therefore,  if $|\alpha\rangle$ is an eigenvector of $H$ with energy $E_\alpha$, 
\begin{equation}H|\alpha\rangle=E_\alpha|\alpha\rangle\,,\end{equation}
then $\brd|\alpha\rangle$ and $b|\alpha\rangle$ are, respectively, eigenvectors with eigenvalues $E_\alpha\pm\lambda$, 
{\it provided} $\brd|\alpha\rangle$ and $b|\alpha\rangle$ are {\it non zero}: 
\begin{eqnarray} 
	H \brd|\alpha\rangle&=&(\brd H+\lambda\brd)|\alpha\rangle=
	(E_\alpha+\lambda)\brd|\alpha\rangle,\label{ha1}\\
	Hb|\alpha\rangle&=&(bH-\lambda b)|\alpha\rangle=(E_\alpha-\lambda)b|\alpha\rangle\,.
	\label{ha2}
 \end{eqnarray}
As in the standard case, these operators then allow one to move along the spectrum, {\it even if it is continuous}, 
as discussed in sec.\  \ref{sec:continuum}. 

The case where ${\cal M}{\cal H}$ is {\it nondiagonalizable}  corresponds here to ${\cal H}$ of rank $1$, and hence to an operator 
$H$ which is just the square of a linear combination of $a$ and $a^\dagger$:
\begin{equation}H_{nd}=(\sqrt{B_-}\,a\pm\sqrt{B_+}\,a^\dagger)^2/2\,.\label{Hnd}\end{equation}
Such $H$ leads to $A=\pm\sqrt{B_+ B_-}$ and $\lambda=0$. This case, 
which includes the free particle case $H\propto P^2$, will be discussed in sec.\  \ref{sec:nondiag}. 

\subsection{The harmonic case\label{B}}
\begin{figure}[h]
	\centering
	\includegraphics[width=0.48\textwidth]{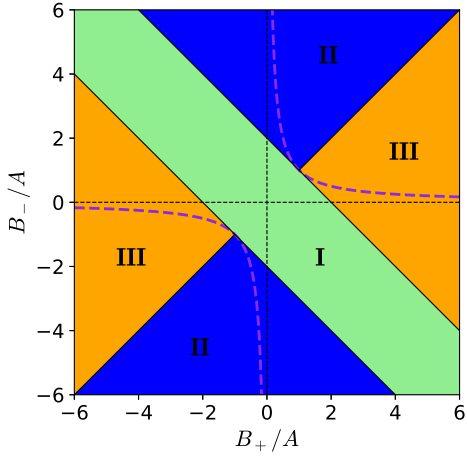}
	\caption{Regions of distinct spectrum for the operator (\ref{H1}) in the case of $B_{\pm}$ real (and $A>0$). 
		{\bf I} denotes the region 
		with discrete positive spectrum (Eq.\ (\ref{cond0})),  {\bf II} that with continuous  complex twofold degenerate spectrum 
		(Eq.\ (\ref{DC})) and {\bf III} that with no convergent eigenfunctions (Eq.\ (\ref{EN})). 
		The dashed curves depict the set of points where ${\cal MH}$ 
		is nondiagonalizable. The hermitian case corresponds  to the line $B_-=B_+$. }
	\label{fig1}
\end{figure}

\begin{figure}[h]
	\centering
	\includegraphics[width=0.48\textwidth]{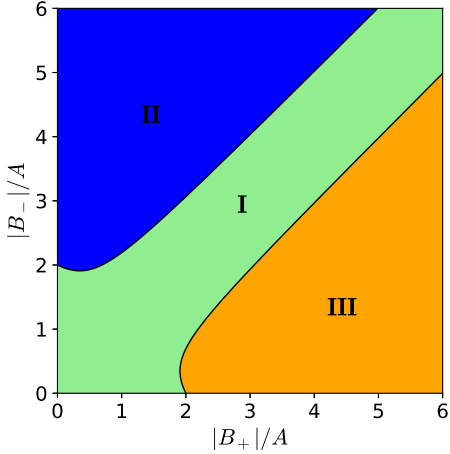}
	\caption{Regions of distinct spectrum for the operator (\ref{H1}) with  complex $B_{\pm}=|B_\pm|e^{i\theta}$ 
		and $\theta=\pi/6$. 
		Same details as Fig.\ \ref{fig1}: In {\bf I}, $H$ has a discrete complex spectrum, 
		while in {\bf II} it has a continuous complex spectrum and in {\bf III} no convergent eigenfunctions. 
		The dotted segment $|B_+|+|B_-|=2A$ indicates the upper limit  of region {\bf I} for $\theta=0$ ($B_{\pm}$ real 
		and positive) whereas dotted lines $|B_+|-|B_-|=\pm 2A$ indicate the border of {\bf I} for $\theta=\pi/2$ ($B_{\pm}$ 
		imaginary); For general $\theta\in(0,\pi/2]$ and $|B_{\pm}|\gg A$, {\bf I} is limited by lines $|B_+|-|B_-|=\pm 2A\sin\theta$. }
	\label{fig2}
\end{figure}

Let $\vaca$ be the vacuum of $a$,  $a\vaca = 0$, and let us assume a vacuum $\vacb$ {\it exists} such that $b\vacb=0$. 
Then, $|0_b\rangle$ is necessarily a gaussian state of the form \cite{schuck:book,*blaizot:book}
\begin{equation}
		\vacb \propto \exp\left(-\frac{v}{2u} a^{\dagger\,2}\right) \vaca \\
		= \sum_{n=0}^{\infty} \left( -\frac{v}{2u} \right)^n \frac{\sqrt{(2n)!}}{n!}\Ket{2n_a}\,. 
	\label{eqn:vacb}
\end{equation}
Recalling that $\sum_{n=0}^\infty (\frac{z}{4})^n\frac{2n!}{(n!)^2}$ converges 
to $\frac{1}{\sqrt{1-z}}$  iff 
$|z|\leq 1$ and $z\neq 1$ \footnote{it conditionally converges for $|z|=1$ if  $z\neq 1$, as ensured by Dirichlet 
	criterion: $\sum_{n}^\infty a_n b_n$ converges if $\lim\limits_{n\rightarrow\infty}b_n=0$ and 
	$|\sum_n^k a_n|\leq M\,\forall k$ ($b_n=\frac{(2n)!}{4^n(n!)^2}\approx\frac{1}{\sqrt{\pi n}}$ 
	for large $n$.)} we see that $|0_b\rangle$ has a finite standard norm 
$\langle 0_b|0_b\rangle$ only if $|v|<|u|$, implying 
\begin{equation}\frac{|B_+|}{|B_-|}<\left|\frac{A+\lambda}{A-\lambda}\right|
\,.\label{n0b}\end{equation}
Eq.\ (\ref{n0b}) imposes an upper bound on $|B_+/B_-|$ for given values of $A$ and $B_+B_-$. 
Similarly, assuming a vacuum $|0_{\bar{b}}\rangle$ exists such that  $\bar{b}\,|0_{\bar{b}}\rangle=0$, then 
\begin{equation}
		|0_{\bar{b}}\rangle\propto\exp\left( -\frac{\vrx}{2\bar{u}} a^{\dagger\,^2} \right)\,|0_a\rangle
		= \sum_{n=0}^{\infty} \left( -\frac{\bar{v}}{2\bar{u}} \right)^n \frac{\sqrt{(2n)!}}{n!}\Ket{2n_a}\,, 
			\label{eqn:vacob}
\end{equation}
with  $\langle \bar{0}_b|\bar{0}_b\rangle$ convergent only if  $|\vrx|<|\bar{u}|$, i.e., 
\begin{equation}\frac{|B_-|}{|B_+|}<\left|\frac{A+\lambda}{A-\lambda}\right|\,.
\label{n0br}\end{equation}
Eqs.\ (\ref{n0b})--(\ref{n0br}) determine a common convergence window 
 \begin{equation}
 \frac{|A-\lambda|}{|A+\lambda|}<\frac{|B_+|}{|B_-|}<\frac{|A+\lambda|}{|A-\lambda|}\,
 \label{cw}\,,
 \end{equation}
 equivalent to $|A-\lambda|<|B_\pm|<|A+\lambda|$, within which both $|0_b\rangle$ and $|0_{\bar{b}}\rangle$ are 
 well defined. For $A\geq 0$, such window can exist only if $A>0$ and ${\rm Re}(\lambda)>0$, which justifies 
 our previous sign  choice of $\lambda$. This window corresponds to region {\bf I} in Figs.\ \ref{fig1}--\ref{fig2}.  
 
On the other hand, their overlap $\langle 0_{\bar{b}}|0_b\rangle$ converges iff
 \begin{equation}
 \left|\frac{v\vrc}{u\urc}\right|=\left|\frac{A-\lambda}{A+\lambda}\right|\leq 1\,,
 \label{des1}
 \end{equation}
and $v\vrc\neq u\urc$, but these conditions are always satisfied due to  Eq.\ (\ref{det1}) 
and the choice ${\rm Re}(\lambda)\geq 0$ (for $A\geq 0$). In particular, if Eq.\ (\ref{cw}) holds, 
Eq.\ (\ref{des1}) is always fulfilled. 
	
It is now natural to define, for $m,n\in{\mathbb{N}}$,  the states 
\begin{equation}
|n_b\rangle=\frac{(\brd)^n}{\sqrt{n!}}|0_b\rangle\,,\;\;\;
|m_{\bar{b}}\rangle=\frac{(b^{\dagger})^m}{\sqrt{m!}}|0_{\bar{b}}\rangle\,,\label{sts}
\end{equation}
which, since  $[\brd b,\brd] =\brd$  and $[b^\dagger\bar{b},b^\dagger]=b^\dagger$,  satisfy 
\begin{equation}
\brd b|n_b\rangle=n|n_b\rangle\,,\;\;\;b^\dagger\bar{b}|m_{\bar{b}}\rangle=m|m_{\bar{b}}\rangle\,,
\label{sts2}
\end{equation}
with 
\begin{equation}
\Braket{m_{\bar{b}}|n_b} = \delta_{mn}\langle 0_{\bar{b}}|0_b\rangle,
\label{eqn:biortonormal}
\end{equation}
implying that $\{|n_b\rangle\}$ and $\{\Ket{n_{\bar{b}}}\}$ form a {\it biorthogonal set} \cite{brody:2013:jpa}. 
Adding ``normalization'' factors $u^{-1/2}$ and $\bar{u}^{-1/2}$ in  
(\ref{eqn:vacb})--(\ref{eqn:vacob})  
directly leads to  $\langle 0_b|0_{\bar{b}}\rangle=1$. 
Note, however,  that the $|n_b\rangle$ are not orthogonal among themselves, nor are the  $|m_{\bar{b}}\rangle$. 
Since $\bar{b}^\dagger=C^{-1}[b^\dagger+(u\bar{v}-v\bar{u})^*b]$, with 
$C=|u|^2-|v|^2=[b,b^\dagger]$, the $|n_b\rangle$ are linear combinations of standard Fock states 
$\propto (b^\dagger)^k|0_b\rangle$ with $k=n,n-2\ldots$.  Similar considerations hold for the  $|m_{\bar{b}}\rangle$. 

We can then write, in agreement with Eqs.\ (\ref{ha1})--(\ref{ha2}), 
\begin{equation}
H \Ket{n_b} = \lambda\left(n + \frac{1}{2}\right)\Ket{n_b}, 
\label{eqn:autovalores}
\end{equation}
and also, 
\begin{equation}H^\dagger \Ket{m_{\bar{b}}} = \lambda^*\left(m + \frac{1}{2}\right) \Ket{m_{\bar{b}}}\,,
	\label{Hdag}
\end{equation}
where $H^\dagger=\lambda^*(b^\dagger\bar{b}+\frac{1}{2})$. Hence, in the interval 
(\ref{cw}) there is a lower-bounded {\it discrete spectrum} of both $H$ and $H^\dagger$, as corroborated in 
section \ref{sec:continuum}. 

This discrete spectrum will be proportional to $\lambda$.  Assuming $A$ real, 
$\lambda$ is real and nonzero iff $B_+B_-$ is real and satisfies 
\begin{equation}
B_+B_-< A^2\,.\label{ineq}
\end{equation}
For equal phases of $B_{\pm}$,  it then comprises two cases:\\ i) $B_\pm$ real  ($\theta=0,\pi$) satisfying (\ref{ineq}), 
in which case $\lambda=\sqrt{A^2-|B_+B_-|}<A$ and $u,v,\bar{v}$ in Eq.\ (\ref{eqn:parametros}) are real. 
Here $H$ is invariant under time reversal, since $\mathcal{T}a\mathcal{T} = a$ and $\mathcal{T}a^\dagger\mathcal{T} = a^\dagger$. 
This is the Swanson case \cite{swanson:2004:jmp}. \\
ii)  $B_\pm$ imaginary ($\theta=\pm\pi/2$), in which case $\lambda=\sqrt{A^2+|B_+B_-|}>A$, with $u$ real and $v,\vrc$ imaginary. 
Here $H$ has the antiunitary (or generalized ${\cal PT}$) symmetry  
\cite{wigner:1960:jmp,bender:2010:pla,fernandez:2014:ap,*amore:2014:ap,*amore:2015:ap,*fernandez:2015:ap} $U\mathcal{T}$, 
 with $U$ the phase transformation $(a,a^\dagger) \rightarrow (-ia, ia^\dagger)$. 

For $\lambda$ real, Eq.\  (\ref{cw}) implies $|B_++B_-|<2A$ in case i) and $|B_+-B_-|<2A$ in case ii), 
which can be summarized, for any case with real $\lambda$, as 
\begin{equation}
|B_++B_-^*|<2A\,.\label{cond0}
\end{equation} 
Eq.\ (\ref{cond0}) is equivalent to ${\cal H}+{\cal H}^\dagger$ {\it positive definite}, 
i.e.,
\begin{equation}
{\cal H}+{\cal H}^\dagger>0\,,\label{cond1}\end{equation}
such that  ${\rm Re}[Z^\dagger{\cal H}Z]>0$ $\forall$ $Z=(z_1,z_2)^T\neq 0$. Therefore, {\it both $H$ and $H^\dagger$ 
	 will exhibit a discrete real positive spectrum iff Eq.\ (\ref{cond1}) holds.} Eq. \eqref{cond0} then leads 
 to region {\bf I} in Fig. \ref{fig1}, i.e., the stripe $|B_+ + B_-|\leq 2A$ when $B_\pm$ are real.

On the other hand,  when $\lambda$ is complex the spectrum of $H$ can be made real just by multiplying $H$ by a phase $\lambda^*/|\lambda|$, 
as seen from (\ref{eqn:autovalores}). The ensuing operator $H'$ 
 has the antiunitary symmetry $U{\cal T}$, with 
$U$ the Bogoliubov transformation $(^a_{a^\dagger})\rightarrow U(^a_{a^\dagger})U^{-1}=({\cal W}^*)^{-1}{\cal W}(^a_{a^\dagger})$. 
For complex $\lambda$, the stable sector adopts the form depicted in Fig. \ref{fig2} (sector {\bf I}).
For a common phase $\theta = 0$ ($B_\pm$ real and positive) it is just the triangle $|B_+| + |B_-| < 2A$, 
while for $\theta = \pi/2$ ($B_\pm$ imaginary, equivalent through a phase transformation to $B_{\pm}$ real 
with opposite signs) it corresponds to $||B_+| - |B_-|| < 2A$ (sectors delimited by dotted lines). 
The union of these two sectors leads to the stripe of Fig. \ref{fig1} for $B_\pm$ arbitrary real numbers.
For intermediate phases the stable region is essentially the union of the previous triangle with a 
narrower stripe, asymptotically delimited by the lines $||B_+|-|B_-|| = 2A\,\sin \theta$ for $|B_\pm| \gg A$.
A similar type of diagram for a non-quadratic system was provided in \cite{scholtz:1992:aop}.

	\subsection{The coordinate representation}
We now turn to the representation of $H$ and its eigenstates in terms of coordinate and momentum operators 
\begin{equation}Q=\frac{a+a^\dagger}{\sqrt{2}},\;\;  P=\frac{a-a^\dagger}{i\sqrt{2}}\,,
\end{equation}
 satisfying $[Q,P]=i$. The Hamiltonian (\ref{eqn:H}) becomes 
\begin{eqnarray}
H &=& \frac{1}{2}\left[ \tilde{A}_- P^2 + \tilde{A}_+ Q^2+\tilde{B}\left(QP + PQ\right)\right]\\
&=&\begin{pmatrix}Q&P\end{pmatrix}\tilde{\cal H}\begin{pmatrix}Q\\P\end{pmatrix},
\; \tilde{\cal H}={\cal S}^\dagger{\cal H}{\cal S}=\begin{pmatrix}\tilde{A}_+&\tilde{B}\\\tilde{B}&\tilde{A}_-
\end{pmatrix},
\label{eqn:Hc}
\end{eqnarray}
where ${\cal S}=(^{1\;\;\;i}_{1\,-i})/\sqrt{2}$ and 
\begin{equation}
\tilde{A}_{\pm} = A \pm \frac{B_++B_-}{2}, \quad \tilde{B} = \frac{B_+-B_-}{2i}\,. 
\label{eqn:param2}
\end{equation}
The hermitian case corresponds to $\tilde{A}_{\pm}$ and $\tilde{B}$ real, 
while the generalized discrete positive spectrum case (\ref{cond1}) to 
$\tilde{\cal H} + \tilde{\cal H}^\dagger>0$. Thus, for $B_{\pm}$ real the border $|B_++B_-^*|=2A$ corresponds 
   to $\tilde{A}_-=0$ or $\tilde{A}_+=0$, i.e.\ infinite mass or no quadratic potential,  while for $B_{\pm}$ imaginary to $|\tilde{B}|=A$.

  The diagonal form (\ref{eqn:H_diag}) can then be rewritten as 
 \begin{equation}
  H = \frac{\lambda}{2}\left( P^{\prime\,2} + Q^{\prime\,2} \right), 
  \end{equation}
 where $Q'=\frac{b+\bar{b}^\dagger}{\sqrt{2}}$ and  $P'=\frac{b-\bar{b}^\dagger}{i\sqrt{2}}$ satisfy $[Q',P']=i$ 
 but  are in general no longer hermitian. 
 They are related to $Q,P$ through a general canonical transformation 
\begin{equation}
\begin{pmatrix}Q'\\P'\end{pmatrix}=\tilde{\cal W}\begin{pmatrix}Q\\P
\end{pmatrix}\,,\;\;\;\tilde{\cal W}={\cal S}^\dagger {\cal W}{\cal S}=
\begin{pmatrix}\frac{\alpha+\bar{\alpha}^*}{2}&-\frac{\beta-\bar{\beta}^*}{2i}
\\\frac{\alpha-\bar{\alpha}^*}{2i}&\frac{\beta+\bar{\beta}^*}{2}\end{pmatrix}\,,
\end{equation}
where $(^\alpha_\beta)=u\pm v$,  $(^{\bar{\alpha}}_{\bar{\beta}})=\bar{u}\pm \bar{v}$  and ${\rm Det}(\tilde{\cal W})=1$. 
 Here $\lambda$ can be expressed as 
\begin{equation}\lambda=\sqrt{\tilde{A}_+\tilde{A}_--\tilde{B}^2}\,,\label{laq}\end{equation}
with $\pm\lambda$ the eigenvalues of $\tilde{\cal M}\tilde{\cal H}={\cal S}^\dagger{\cal M}{\cal H}{\cal S}$. 

Setting  $Q|x\rangle=x|x\rangle$, 
the coordinate representations $\psi_0^b(x)\equiv \langle x|0_b\rangle$, 
 $\psi_0^{\bar{b}}(x)\equiv\langle x|0_{\bar{b}}\rangle$ of the vacua can be found from Eqs.\ (\ref{eqn:vacb}) and 
(\ref{eqn:vacob}). They can also be derived by solving the corresponding differential equations 
$\langle x|b|0_b\rangle=0$, 
$\langle x|\bar{b}|0_{\bar{b}}\rangle=0$, i.e., 
\begin{equation}
		\left[\alpha x+\beta \partial_x\right] \psi_0^b(x)=0,\;\;
		\left[\bar{\alpha}x+\bar{\beta}\partial_x\right]\psi_0^{\bar{b}}(x)=0\,,
\end{equation}
and read 
	\begin{eqnarray}
			\psi_0^b(x)&\propto&
			\exp\left[ -\frac{\alpha}{2\beta}x^2 \right]\label{eqn:vacio},\;\;
			\psi_0^{\bar{b}}(x) \propto 
			\exp \left[  -\frac{\bar{\alpha}}{2\bar{\beta}}x^2 \right]\,.
			\label{eqn:autof1}
	\end{eqnarray}
Since ${\rm Re}[\frac{z_1+z_2}{z_1-z_2}]=\frac{|z_1|^2-|z_2|^2}{|z_1-z_2|^2}$ $\forall$ $z_1\neq z_2\in\mathbb{C}$,  
it is verified that they have finite standard norms 
iff $|v|<|u|$ and $|\bar{v}|<|\bar{u}|$.  
	The wave functions of the excited states $\Ket{n_b}$ and $\Ket{m_{\bar{b}}}$ can be similarly obtained by 
	applying  $\brd$ and $b^\dagger$ to the functions (\ref{eqn:vacio}),  according to Eq.\ \eqref{sts}:
	\begin{eqnarray}
			\psi_n^b(x)& =& \frac{1}{\sqrt{n!}} 
			\left[\sqrt{\frac{\bar{\beta}^*}{2\beta}}\right]^n H_n\left( \frac{x}{\gamma} \right)
			\psi_0^b(x)\label{eqn:excitadosb},\\
			\psi_m^{\bar{b}}(x)& =& 
			\frac{1}{\sqrt{m!}}\left[\sqrt{\frac{\beta^*}{2\bar{\beta}}}\right]^m H_m\left( \frac{x}{\gamma^*} \right)
			\psi_0^{\bar{b}}(x)\label{eqn:excitadosbr},
	\end{eqnarray}
	where $\gamma = \sqrt{\beta\bar{\beta}^*}$ and $H_n(x)$ is the Hermite polynomial of degree $n$. 
	These functions  satisfy the biorthogonality relation \eqref{eqn:biortonormal}, i.e., 
	$\int_{-\infty}^{\infty} \psi_m^{\bar{b} *}(x) \psi_n^b(x) \mathrm{d}x = \delta_{mn}\Braket{0_{\bar{b}}|0_b}$, 
	with $\Braket{0_{\bar{b}}|0_b} = 1$ if normalization factors $(\sqrt{\pi}\beta)^{-1/2}$ and $(\sqrt{\pi}\bar{\beta})^{-1/2}$ are 
	added in \eqref{eqn:autof1}. 
	They are verified to be the finite norm solutions to the Schr\"{o}dinger equations associated with $H$ and $H^\dagger$  respectively. 
	In the case of $\psi^b_n(x)$, the latter reads 
		\begin{equation}
	-\frac{1}{2}\tilde{A}_- \psi^{\prime\prime} - i \tilde{B} \left [ x \psi^\prime + \frac{\psi}{2} \right] +
	 \frac{1}{2}\tilde{A}_+ x^2 \psi = E\psi\,,
	\label{eqn:schrodinger}
\end{equation}
	 with $E = \lambda(n+1/2)$,  
	 while in the case of $\psi_m^{\bar{b}}(x)$,  $\tilde{A}_\pm$,  $\tilde{B}$ are to be replaced by $\tilde{A}_\pm^*$  
	 and $\tilde{B}^*$, with $E=\lambda^*(m+1/2)$.  

\subsection{The case of continuous spectrum  \label{sec:continuum}}
If $|v/u| < 1$ but $|\vrx/\bar{u}| > 1$, the vacuum $|0_{\bar{b}}\rangle$ of 
$\bar{b}$ is no longer well defined, since the coefficients of its expansion in the states $|n_a\rangle$, Eq.\ (\ref{eqn:vacob}), 
become increasingly large for large $n$, and the associated eigenfunction $\psi_0^{\bar{b}}(x)$, Eq.\ (\ref{eqn:vacio}), 
becomes divergent. This situation occurs whenever 
\begin{equation}
\frac{|B_+|}{|B_-|}<\frac{|A-\lambda|}{|A+\lambda|}\,,\label{DC}
\end{equation} 
i.e.\  below the window \eqref{cw}, and corresponds to regions {\bf II} in Figs.\  \ref{fig1} and \ref{fig2}. 
The same occurs with the excited states $\Ket{n_{\bar{b}}}$ defined in Eq.\  \eqref{sts}.

Instead, it is now the operator $\brd$ which has a convergent vacuum, namely 
\begin{equation}
	\Ket{0_{\brd}} \propto\sum_{n=0}^{\infty} \left( -\frac{\bar{u}^*}{2\vrx^*} \right)^n \frac{\sqrt{2n!}}{n!} \Ket{2n_a},
	\label{eqn:vacio_brd}
\end{equation}
satisfying $\brd\Ket{0_{\brd}} = 0$.
Since we can write $H$ as 
\begin{equation}
	H = -\lambda\,[(-b) \brd + 1/2],
\end{equation}
it becomes clear that $H\Ket{0_{\brd}} = -\lambda/2\,\Ket{0_{\brd}}$. 
Moreover, due to the commutation relation $[\brd,-b] = 1$, we may as well consider $-b$ as a creation operator 
and $\brd$ as an annihilation operator, and define the states
\begin{equation}
	\Ket{n_{\brd}} =  \frac{(-b)^n\Ket{0_{\brd}}}{\sqrt{n!}},
	\label{eqn:excitados_brd}
\end{equation}
which then satisfy $-b \brd \Ket{n_{\brd}} = n \Ket{n_{\brd}}$, and hence
\begin{equation}
	H \Ket{n_{\brd}} = -\lambda\left(n + \frac{1}{2}\right)\Ket{n_{\brd}}.
\end{equation}

Since the previous states $|0_b\rangle$ and $\Ket{n_b}$ remain convergent, and Eq.\ \eqref{eqn:autovalores} still holds, 
it is seen that $H$ posseses in this case two sets of discrete eigenstates constructed from the vacua of $b$ 
and $\brd$, with {\it opposite energies}. The wave functions of the ``negative'' band are given by
\begin{equation}
	\begin{aligned}
		\psi_0^{\brd}(x) &\propto \exp\left[ \frac{\bar{\alpha}^*}{2\bar{\beta}^*}\,x^2 \right],\\
		\psi_n^{\brd}(x) &\propto \frac{1}{\sqrt{n!}}\left[ \sqrt{\frac{\beta}{2\bar{\beta}^*}} \right]^n H_n\left( \frac{ix}{\gamma} \right)\psi_0^{\brd}(x),
	\end{aligned}
	\label{eqn:estadosbrd}
\end{equation}
which are convergent since now $\Re(\bar{\alpha}^*/\bar{\beta}^*) < 0$.

However, these eigenvalues do not exhaust, remarkably,  the entire spectrum.  
The Schr\"{o}dinger equation \eqref{eqn:schrodinger} has in the present case {\it two} linearly independent bounded  
eigenstates $|\nu_b\rangle$ and $|\nu_{\bar{b}^\dagger}\rangle$,  for {\it any complex energy} 
\begin{equation}
	E_{\nu} = \lambda\left( \nu+\frac{1}{2} \right),
\end{equation}
with $\nu \in \mathbb{C}$. As demonstrated in the appendix, the associated  eigenfunctions 
$\psi_\nu^b(x) = \Braket{x|\nu_b}$ and $\psi_\nu^{\brd}(x) = \Braket{x|\nu_{\brd}}$ are given explicitly by:
\begin{widetext}
\begin{eqnarray}
		\psi_{\nu}^b(x) &=& 
		\Xi(\nu)\left( \sqrt{\frac{\brc}{2\beta}} \right)^n\exp\left( -\frac{i\tilde{B} + \lambda}{2\tilde{A}_-}\, x^2 \right)
		\left[ H_{\nu}\left( \frac{x}{\gamma} \right) + (-1)^n H_{\nu}\left( -\frac{x}{\gamma} \right) \right],
		\label{eqn:solu1gral}\\
		\psi_{\nu}^{\brd}(x) &=& 
		\Xi(\nu)\left( \sqrt{\frac{\beta}{2\brc}} \right)^n\exp\left( -\frac{i\tilde{B} - \lambda}{2\tilde{A}_-}\,x^2 \right) 
		\left[ H_{\nu}\left( \frac{ix}{\gamma} \right) + (-1)^n H_{\nu}\left( -\frac{ix}{\gamma} \right) \right],
		\label{eqn:solu2gral}
\end{eqnarray}
\end{widetext}
where  $n = \lfloor \Re(\nu) \rfloor$, with $\lfloor x \rfloor$  
 the greatest integer lower than $x$ (floor function), and
\begin{equation}
	\Xi(\nu) = \begin{cases}
		\sqrt{(|\nu|-1)!} &\quad \nu = -1,-2,\ldots\\
		\frac{1}{\sqrt{\Gamma(\nu+1)}} &\quad \mbox{otherwise}
	\end{cases}\,.
\end{equation}
 For integer $\nu \geq 0$, these functions are proportional to the previous expressions  \eqref{eqn:excitadosb} 
and \eqref{eqn:estadosbrd}. For general $\nu \in \mathbb{C}$, they satisfy
\begin{eqnarray}
	H \Ket{\nu_b}& =& \lambda \left( \nu + \frac{1}{2} \right)\Ket{\nu_b},\label{eqn:II-1}\\
	H \ket{\nu_{\brd}}& =& -\lambda\left( \nu + \frac{1}{2} \right) \Ket{\nu_{\brd}}\,,
\end{eqnarray}
with 
\begin{equation}
\begin{aligned}
b\Ket{\nu_b} &\propto \sqrt{\nu}\Ket{\nu-1_{b}},\\
\brd\Ket{\nu_b} &\propto \left\{\begin{array}{lcl}\sqrt{\nu+1} \Ket{\nu+1_b} && (\nu \neq -1)\\
\Ket{0_b}&&(\nu=-1)\end{array}\right.\,,
\end{aligned}
\end{equation}
\begin{equation}
\begin{aligned}
\brd\Ket{\nu_{\brd}} &\propto \sqrt{\nu}\Ket{\nu-1_{\brd}},\\
(-b)\Ket{\nu_{\brd}} &\propto \left\{\begin{array}{lcl}\sqrt{\nu+1} \Ket{\nu+1_{\brd}} && (\nu \neq -1)\\\Ket{0_{\brd}}&&(\nu=-1)\end{array}\right.\,,
\label{eqn:II-6}
\end{aligned}
\end{equation}
where the proportionality constant is a phase factor. 
Expressions \eqref{eqn:II-1}--\eqref{eqn:II-6} are in agreement with Eqs.\  \eqref{ha1}--\eqref{ha2}. 
They are valid in this region for both real or complex $\lambda$. 

Note that if $\brd\Ket{\minus 1_b}$ would vanish, then $\Ket{\minus 1_b}$ would be proportional to $\Ket{0_{\brd}}$, which is not the case.
A similar argument holds for $b\Ket{\minus 1_{\brd}}$.
It is also verified that in the case of discrete spectrum (region {\bf I}), 
such state $\Ket{\minus 1_b}$ does not exist, i.e., 
the solution of the first order differential equation $\Braket{x|\brd|\minus 1_b} = \Braket{x|0_b}$ is divergent.
In addition, we remark that Eqs. \eqref{eqn:solu1gral} and \eqref{eqn:solu2gral} are always linearly independent 
solutions of the Schr\"{o}dinger equation \eqref{eqn:schrodinger}, 
but  in region {\bf I} the function  \eqref{eqn:solu2gral} is always divergent whereas \eqref{eqn:solu1gral} is 
divergent except for $\nu = n = 0,1,2,\ldots$. 

\subsection{The case of  no convergent eigenstates}
If now $|\vrx/\bar{u}| < 1$ but  $|v/u| > 1$, i.e., 
\begin{equation}\frac{|B_+|}{|B_-|} >\frac{|A+\lambda|}{|A-\lambda|}\,,\label{EN}
\end{equation} 
neither $b$ nor $\bar{b}$ have a convergent vacuum, so that  the eigenstates $\Ket{n_b}$ and $\Ket{n_{\brd}}$ of sec.\ \ref{B} are not well defined.
In fact, Eqs. \eqref{eqn:solu1gral} and \eqref{eqn:solu2gral} become {\it divergent for \emph{any} $\nu$}, so that $H$ has no convergent 
eigenfunctions for {\it any} value of $E$. This case corresponds to regions {\bf III} in Figs.\ \ref{fig1}--\ref{fig2}.  

On the other hand, it is the operator $b^\dagger$ which now has a well defined vacuum $\Ket{0_{\bd}}$, in addition to $\bar{b}$, 
which preserves its vacuum $|0_{\bar{b}}\rangle$. 
Therefore, one can define the states $\Ket{n_{\bd}}$ and $\Ket{n_{\bar{b}}}$ in the same way as the treatment of previous section, 
and also $\Ket{\nu_{\bd}}$ and $\Ket{\nu_{\bar{b}}}$ for any $\nu\in\mathbb{C}$, which will be eigenstates of $H^\dagger$. 
Hence, in this case $H^\dagger$, rather than $H$,  has two linearly independent bounded  eigenfunctions for every 
complex value of $E$. In contrast, in ${\bf II}$ $H^\dagger$ has no bounded eigenstate. 

\subsection{Non diagonalizable case \label{sec:nondiag}}
The matrix ${\cal M H}$ becomes non diagonalizable when $\lambda = 0$, i.e.\ ${\rm rank}\,{\cal H}=1$. This case occurs whenever $B_+ B_- = A^2$  
and corresponds to the dashed curve in Fig.\ \ref{fig1}, which lies in regions {\bf II} and {\bf III}. 
The operator $H$ takes here the single square form (\ref{Hnd}). 

We first  analyze the sector lying in region {\bf II}. In the limit $B_+ \rightarrow 0$, 
with $B_- = A^2/B_+ \rightarrow \infty$, 
$H$ becomes proportional to  $a^2$. Its eigenstates then become the well known {\it coherent states}
\begin{equation}|\alpha_a\rangle\propto\exp[\alpha a^\dagger]|0_a\rangle\,, \end{equation}
satisfying $a\Ket{\alpha_a} = \alpha\Ket{\alpha_a}$, $\alpha \in \mathbb{C}$, with $\frac{2B_+}{A^2}H \Ket{\pm\alpha_a}\rightarrow \alpha^2 \Ket{\pm \alpha_a}$. 
This implies a {\it continuous two-fold degenerate spectrum},  as in the rest  of region {\bf II}.  
The spectrum of $H$ in {\bf II} is then similar to that of $a^2$, reflecting the fact that here both $b$ and $\brd$ have a convergent 
vacuum and are then annihilation operators. 

In fact, for $\lambda\rightarrow 0$ and $A>0$, the operators $b$ and $\brd$ of Eq.\ (\ref{eqn:bogoliubov}) become \emph{proportional}, 
i.e. $\brd \rightarrow \sqrt{B_-/B_+}\,b$, such that $H\propto b^2$ at leading order. At the curve $\lambda=0$ and within region {\bf II}, $H$ takes the exact form  
\begin{equation}
	H = {\textstyle\frac{|B_-|-|B_+|}{2}\,\tilde{b}^2,\;\;\;\; \tilde{b} = \frac{\sqrt{B_-}\,a + \sqrt{B_+}\,a^\dagger}{\sqrt{|B_-|-|B_+|}}},
\end{equation}
where $\tilde{b}$ fulfills $[\tilde{b},\tilde{b}^\dagger]=1$ and has  a \emph{convergent} vacuum $\Ket{0_{\tilde{b}}}$ since here $|B_+|<|B_-|$. 
It then represents a proper  {\it annihilation} operator. 
The eigenstates of $H$ become its coherent states $\Ket{\alpha_{\tilde{b}}}\propto \exp[\alpha \tilde{b}^\dagger]\Ket{0_{\tilde{b}}}$ 
satisfying $\tilde{b}\Ket{\alpha_{\tilde{b}}} = \alpha\Ket{\alpha_{\tilde{b}}}$, such that 
\begin{equation}
	H\Ket{\pm\,\alpha_{\tilde{b}}} = {\textstyle\frac{|B_-|-|B_+|}{2}}\,\alpha^2\Ket{\pm\,\alpha_{\tilde{b}}}\,,
\end{equation}
with $\alpha \in \mathbb{C}$. The spectrum is then complex continuous and two-fold degenerate, as in the rest of sector  {\bf II}. The eigenfunctions become  
\begin{equation}
	\psi_\alpha(x) = \Braket{x|\alpha_{\tilde{b}}} \propto e^{-\frac{1}{2}\frac{\sqrt{B_-}+\sqrt{B_+}}{\sqrt{B_-}-\sqrt{B_+}} 
		\left( x-\sqrt{2}\alpha\frac{\sqrt{|B_-|-|B_+|}}{\sqrt{B_-}+\sqrt{B_+}} \right)^2 }.
\end{equation}

On the other hand, in region {\bf III}, $|B_+| > |B_-|$ and along the curve $\lambda=0$ we have instead 
\begin{equation}
H = {\textstyle\frac{|B_+|-|B_-|}{2}\,\tilde{b}^{\dagger\,2},\;\;\;\; \tilde{b}^\dagger = \frac{\sqrt{B_-}\,a + \sqrt{B_+}\,a^\dagger}{\sqrt{|B_+|-|B_-|}}},
\end{equation}
with  $\tilde{b}^\dagger$  a proper {\it creation} operator  
satisfying $[\tilde{b},\tilde{b}^\dagger]=1$ and having no bounded vacuum. Hence, here $H$ has  no bounded eigenstates while $H^\dagger$ has has a continuous complex spectrum. 

Finally, in the hermitian limit $|B_+|=|B_-|=A$, i.e.\ when the curve $\lambda=0$ crosses the border between {\bf II} and {\bf III}, 
$H\rightarrow\frac{A}{2}(e^{-i\phi} a+e^{i\phi}a^\dagger)^2$, becoming proportional to $Q^2$ (or equivalently, to $P^2$ if $\phi=\pi/2$). It then possesses a continuous two-fold degenerate nonnegative {\it real} spectrum,  although with non 
normalizable eigenstates ($\Ket{x}$ or $\Ket{p}$). This case corresponds in Fig.\ \ref{fig1} to the two ``critical'' 
points where {\it all three regions} {\bf I, II, III} {\it merge},  i.e., $|v/u| = |\bar{v}/\bar{u}| = 1$.  
Thus, at the non-diagonalizable curve $\lambda=0$, $H$ is proportional to  the square of:  an annihilation operator inside region {\bf II},
a creation operator inside region {\bf III}, and a coordinate or momentum operator at the crossing with the Hermitian case.

\subsection{Intermediate regions}

We finally discuss the border between regions {\bf I} and {\bf II} or {\bf III}.
These intermediate lines have either $|v/u| = 1$ or $|\vrx/\bar{u}| =1$.
When crossing from {\bf I} to {\bf II} ({\bf III}), $\bar{b}$ ($b$) undergoes an {\it annihilation $\rightarrow$ creation} transition, 
loosing its bounded vacuum and becoming at the crossing  a coordinate or momentum. 

As can be verified from Eqs.\ \eqref{eqn:solu1gral} and \eqref{eqn:solu2gral} when $\tilde{A}_- \neq 0$, at the border between {\bf I} and {\bf II} $H$ 
has still a discrete spectrum and satisfies  Eq.\ \eqref{eqn:autovalores}, since \eqref{eqn:solu1gral} remains convergent just for $\nu = n$.
On the other hand, \eqref{eqn:solu2gral} has no longer a finite norm since $(i\tilde{B}-\lambda)/(2\tilde{A}_-)$ is an imaginary number. 
However, the dual states $\Ket{0_{\bar{b}}}$ and $\Ket{n_{\bar{b}}}$, while also lacking a finite norm 
$\Braket{n_{\bar{b}}|n_{\bar{b}}}$, still have {\it finite} biorthogonal  norms $\Braket{m_{\bar{b}}|n_b}$, fulfilling Eq.\ (\ref{eqn:biortonormal}).
In contrast, at the border {\bf I}-{\bf III} $H$ ceases to have convergent eigenfunctions for any value of $\nu$, since $\Ket{n_b}$ stops being convergent, 
while dual states $\Ket{n_{\bar{b}}}$ remain convergent.

When $\tilde{A}_- = 0$, which corresponds to the case $B_\pm$ real and $B_+ + B_- = 2A$ (the border between {\bf I} 
and regions {\bf II}--{\bf III} in Fig.\ \ref{fig1}), we have $\vrx =\bar{u}$.
In this case, and for $A \neq B_-$, Eq. \eqref{eqn:schrodinger} becomes of first order  and has a unique solution given by 
\begin{equation}
	\psi_\nu^b(x) \propto e^{-\frac{Ax^2}{2(B_--A)}} x^\nu,
	\label{eqn:psi_borde}
\end{equation}
where we have set $E = \lambda(\nu+1/2)$, with $\lambda = B_- - A$, along this line.
Hence, at the border with region {\bf III} ($B_- < A$) Eq. \eqref{eqn:psi_borde} is always divergent for $|x| \rightarrow \infty$, while  at the border 
with {\bf II} it is always convergent for $|x| \rightarrow \infty$ yet regular at $x = 0$ {\it just  for} $\nu = n = 0,1,2,\ldots$, 
as in the previous case. For these values, Eq. \eqref{eqn:psi_borde} becomes proportional to Eq. \eqref{eqn:excitadosb}.

Regarding the dual states, at this line $\bar{b} = \brd = \sqrt{2}\bar{u} Q$, (since $\bar{u}$ is real) 
and as such $\Ket{0_{\bar{b}}}$ is the state with $Q = 0$, i.e., $\Braket{x|0_{\bar{b}}} \propto \delta(x)$.
In fact, for $\vrx \rightarrow \bar{u}$ the coordinate representation of the state $\Ket{0_{\bar{b}}}$ in \eqref{eqn:vacob} becomes a delta function, 
as also seen from Eq.\ (\ref{eqn:vacio}):
	\begin{equation}
		\Braket{x|0_{\bar{b}}} \rightarrow \frac{e^{-x^2/2}}{\pi^{1/4}} \sum_{n=0}^{\infty} \frac{H_{2n}(x)H_{2n}(0)}{2^{2n} (2n)!} = \pi^{1/4} \delta(x),
	\label{eqn:delta_exp}
\end{equation}
where we have used $\delta(x) = \Braket{x|0_{\bar{b}}} = \sum_{n=0}^{\infty} \langle x | n_a \rangle \langle n_a | 0_{\bar{b}} \rangle$.
It is then still verified that $\Braket{0_{\bar{b}}|0_b}$ is a finite number.
The same holds for the remaining states $\Ket{n_{\bar{b}}}$, with $\Braket{x|n_{\bar{b}}}$ involving derivatives of the delta function, 
such that Eq.\ (\ref{eqn:biortonormal}) still holds. 

\section{The general $N$-dimensional case}
We now discuss the main features of the $N$-dimensional case. We consider a general $N$-dimensional quadratic form in 
boson operators $a_i$, $a^\dagger_j$ satisfying $[a_i,a^\dagger_j]=\delta_{ij}$, $[a_i,a_j]=0$,
$i,j=1,\ldots,N$: 
\begin{eqnarray}
	H&=&\sum_{i,j}A_{ij}a^\dagger_i a_j+\frac{1}{2}(B^+_{ij}a^\dagger_ia^\dagger_j+B^-_{ij}a_ia_j)\label{hgg0}\\
	&=&\frac{1}{2}\begin{pmatrix}a^\dagger&a\end{pmatrix}{\cal H}\begin{pmatrix}a\\a^\dagger\end{pmatrix}\,,\;\;{\cal H}=\begin{pmatrix}A&B_+\\B_-&A^T
\end{pmatrix}\,.\label{hgg}
\end{eqnarray}
Here $B_{\pm}$ are {\it symmetric} $N\times N$ matrices of elements $B^{\pm}_{ij}$, 
such that  ${\cal H}$  satisfies 
\begin{equation}
	{\cal H}^T={\cal R H R}\,,\;\;{\cal R}=\begin{pmatrix}0&{\mathbbm 1}\\{\mathbbm 1}&0\end{pmatrix}\,. \label{br}
\end{equation}
Following the treatment of \cite{rossignoli:2005:pra_0} for the general hermitian case, 
we define new operators $b_i$,  $\bar{b}^\dagger_i$ through a generalized Bogoliubov transformation 
\begin{equation}\begin{pmatrix}b\\\bar{b}^\dagger\end{pmatrix}={\cal W}\begin{pmatrix}a\\a^\dagger\end{pmatrix},\;
{\cal W}=\begin{pmatrix}U&V\\\bar{V}^*&\bar{U}^*\end{pmatrix}\,,
\end{equation}
where again $\bar{b}^\dagger_i$ may not coincide with $b^\dagger_i$  although the bosonic commutation relations are preserved:
\begin{equation}[b_i,\bar{b}^\dagger_j]=\delta_{ij},\;\;[b_i,b_j]=[\tilde{b}_i^\dagger,
\tilde{b}_j^\dagger]=0. \label{70} \end{equation}
These conditions imply \cite{rossignoli:2005:pra_0,rossignoli:2009:pra_0}
\begin{equation}
{\cal W M R W}^T{\cal R}={\cal M}\,,\label{WS}\end{equation}
(${\cal M}$ is the matrix (\ref{M}) extended to $2N\times 2N$) i.e., 
\begin{eqnarray}
	U(\bar{U}^*)^T-V(\bar{V}^*)^T&=&{\mathbbm 1},
	\label{WS2}\\
	VU^T-UV^T&=&0,\; \bar{V}\bar{U}^T-\bar{U}\bar{V}^T=0\,.
	\label{WS3}
\end{eqnarray}
We can then rewrite $H$ exactly as in Eqs.\ (\ref{Hb0})--(\ref{Hb}):
\begin{eqnarray}
H&=&\frac{1}{2}\begin{pmatrix} \brd&b\end{pmatrix}\mathcal{H}'\begin{pmatrix}b\\\brd\end{pmatrix}\,,\;
\mathcal{H}' = {\cal M}{\cal W}{\cal M}{\cal H}{\cal W}^{-1}\,,\end{eqnarray}
where ${\cal H}'$ has again the form (\ref{hgg}) and satisfies  (\ref{br}) 
due to Eq.\ (\ref{WS}). 
The problem of obtaining a normal mode  representation 
\begin{equation}
H=\sum_i \lambda_i(\bar{b}_i^\dagger b_i+\frac{1}{2})\,,\label{diagg}
\end{equation}
leads then to the diagonalization of the matrix  
\begin{equation}
	{\cal MH} = \begin{pmatrix}
		A & B_+ \\
		-B_- & -A^T
	\end{pmatrix},
	\label{eqn:MHmatriz}
\end{equation}
which is that representing the commutation relations of Eq. \eqref{eqn:conmuta} in the present general case: 
$[H,(^a_{a^\dagger})]={\cal MH} (^a_{a^\dagger})$. 

A basic result is that the eigenvalues of (\ref{eqn:MHmatriz})  {\it always come in pairs of  opposite sign}, 
as in the hermitian case \cite{rossignoli:2005:pra_0} (see also  \cite{fernandez:2016:arxiv}): 
Noting that ${\cal RM}=-{\cal MR}$ and ${\cal M}^2={\cal R}^2={\mathbbm 1}$, Eq.\ (\ref{br}) implies 
\[({\cal M H}-\lambda{\mathbbm 1})^T={\cal RHRM}-\lambda{\mathbbm 1}={\cal RM}({\cal MH}+\lambda{\mathbbm 1}){\cal RM}\]
and hence ${\rm Det}[{\cal MH}-\lambda{\mathbbm 1}]={\rm Det}[{\cal MH}+\lambda{\mathbbm 1}]$, 
entailing that if $\lambda$ is an eigenvalue of ${\cal M} {\cal H}$, so is $-\lambda$.  

From Eq.\ (\ref{br}) we also see that if $Z_i$ are eigenvectors of ${\cal MH}$ satisfying 
${\cal M}{\cal H}Z_i=\lambda_i Z_i$,  then $Z_i^T{\cal R M}Z_j(\lambda_i+\lambda_j)=0$, 
implying  the orthogonality relations 
\begin{equation}
	Z_i^T{\cal RM}Z_j=0\;\;\;\;(\lambda_i\neq -\lambda_j)\,.\label{o1}
\end{equation}
The pairs $(b_i, \brd_i)$ emerge then from the eigenvectors $Z_i, Z_{\bar{i}}$ associated to {\it opposite}  eigenvalues  
$\pm \lambda_i$, which are to be scaled such that
\begin{equation}Z_i^T{\cal RM}Z_{\bar{i}}=1\,.\label{o2}\end{equation}
Writing $Z_i = \begin{pmatrix} \bar{U}^* & -\bar{V}^* \end{pmatrix}^T_i$ and $Z_{\bar{i}} = \begin{pmatrix} -V & U\end{pmatrix}^T_i$, we can 
form with them the eigenvector matrix ${\cal W}^{-1}$, with Eqs.\ (\ref{o1})--(\ref{o2}) ensuring that ${\cal W}$ will satisfy Eq.\ \eqref{WS}. 

Therefore,  if ${\cal MH}$ is {\it diagonalizable}, a diagonalizing matrix ${\cal W}$ satisfying \eqref{WS}  
will exist  such that $H$ can be written in the diagonal form (\ref{diagg}). 
The $N$-dimensional $H$  can then be reduced  to a sum 
 of $N$ commuting one-dimensional systems ({\it complex normal modes}) described by operators $H_i=\lambda_i (\bar{b}^\dagger_i b_i+\frac{1}{2})$.  
The normal operators $b_i$, $\bar{b}^\dagger_i$, satisfy 
\begin{equation}
[H,b_i]=-\lambda_i b_i,\;\;[H,\bar{b}^\dagger_i]=\lambda_i \bar{b}^\dagger_i\,,
\end{equation}
diagonalizing the commutator algebra with $H$ and satisfying then Eqs.\ (\ref{ha1})--(\ref{ha2}) 
$\forall$ $b=b_i$. 

Now, if a common vacuum $|0_b\rangle$ exists such that 
\begin{equation}
b_i|0_b\rangle=0\,,
\end{equation}
for $i=1,\ldots,N$, it must necessarily be of the form \cite{schuck:book}
\begin{equation}
|0_b\rangle\propto \exp[-\frac{1}{2}\sum_{i,j}(U^{-1}V)_{ij}a^\dagger_i a^\dagger_j]|0_a\rangle\,,\label{0g}
\end{equation}
where $U^{-1}V$ is a {\it symmetric} matrix due to Eq.\ (\ref{WS3}). Eq.\ (\ref{0g}) can be directly 
checked by application of $b_i$. Similarly, assuming a common vacuum $|0_{\bar{b}}\rangle$ exists such that  
\begin{equation}
\bar{b}_i|0_{\bar{b}}\rangle=0\,,
\end{equation}
for $i=1,\ldots,N$, it must be of the form 
\begin{equation}
|0_{\bar{b}}\rangle\propto \exp[-\frac{1}{2}\sum_{i,j}(\bar{U}^{-1}\bar{V})_{ij}a^\dagger_i a^\dagger_j]|0_a\rangle\,.
\end{equation}
Assuming these series are convergent, which implies that $U^{-1}V$ 
and $\bar{U}^{-1}\bar{V}$ have both  all singular values $\sigma_i<1$, $\bar{\sigma}_i<1$, we can define the states 
\begin{eqnarray}
	|n_1,\ldots,n_N\,_b\rangle&=&\left(\prod_i
	\frac{(\bar{b}^\dagger_i)^{n_i}}{\sqrt{n_i!}}\right)|0_b\rangle,\\
	|m_1,\ldots,m_N\,_{\bar{b}}\rangle&=&\left(\prod_i
	\frac{(b^\dagger_i)^{m_i}}{\sqrt{m_i!}}\right)|0_{\bar{b}}\rangle\,.
\end{eqnarray}
Due to the commutation relations (\ref{70}),  these states form again a biorthogonal set,  
\begin{equation}
\langle m_1,\ldots,m_N\,_{\bar{b}}|n_1,\ldots,n_N\,_b\rangle=\delta_{m_1 n_1}\ldots\delta_{m_N n_N}\langle 0_{\bar{b}}|0_b\rangle\,,
\end{equation}
and satisfy 
\begin{eqnarray}
	H||n_1,\ldots,n_N\,_b\rangle&=&\sum_i \lambda_i\left(n_i+\frac{1}{2}\right)|n_1,\ldots,n_N\,_b\rangle,\\
	H^\dagger|m_1,\ldots,m_N\,_{\bar{b}}\rangle&=&\sum_i \lambda_i^*\left(m_i+\frac{1}{2}\right)|m_1,\ldots,m_N\,_{\bar{b}}\rangle\,.
\end{eqnarray}
Thus, both $H$ and $H^\dagger$ possess in this case a {\it discrete} spectrum. 
Such spectrum can be real if $H$ has some antilinear (generalized ${\cal PT}$) symmetry (for instance, ${\cal H}$ real). 

In a general situation, a common vacuum may exist just for a certain subset of operators $b_i$ and 
$\bar{b}_i$, leading to terms $H_i$ with behaviors similar to those encountered  in the previous section. 
An important difference is to be found in the non-diagonalizable cases: The corresponding modes  may  not  
necessarily be of the form (\ref{Hnd}), and are not necessarily associated with vanishing eigenvalues $\lambda_i=0$,  
since Jordan forms of higher dimension can arise, as was already shown in two-dimensional systems 
\cite{rossignoli:2009:pra_0,rebon:2014:pra_0}, in the context of hermitian yet unstable Hamiltonians. 
Besides, ${\cal MH}$ may remain diagonalizable in the presence of vanishing eigenvalues  \cite{rossignoli:2009:pra_0,colpa:1986:pa}.

\section{Conclusions}
We have first analyzed the spectrum and normal modes of a general one-dimensional quadratic bosonic form,  
showing that it can exhibit three distinct regimes:\\  {\bf i)}  An harmonic phase characterized by a 
discrete spectrum of both $H$ and $H^\dagger$,  with bounded eigenstates constructed from gaussian vacua, 
which form a biorthogonal set. Such phase, which comprises the cases considered in \cite{swanson:2004:jmp,jones:2005:jpa},  
arises when the deviation from the stable hermitian case is not ``too large'' (Eq.\ (\ref{cw}), 
equivalent to (\ref{cond0})--(\ref{cond1}) for $\lambda>0$), in which case the generalized normal 
boson operators $\bar{b}^\dagger$, $b$ can be considered as creation and annihilation operators respectively.   
According to the phase of $\lambda$, the discrete spectrum can be real  or complex,  but in the latter 
it can be made real by applying a trivial phase factor (as opposed to discrete regimes in nonquadratic  Hamiltonians \cite{fernandez:2014:amc}). 

 {\bf ii)} A coherent-like phase where $H$ exhibits a complex twofold degenerate continuous spectrum while $H^\dagger$ 
 has no bounded eigenstates. It corresponds to large deviations from the hermitian harmonic case. The normal operators $\bar{b}^\dagger$, $b$
 can be considered as a pair of annihilation operators, each with a convergent vacuum yet still satisfying a bosonic commutator. 
 The spectrum is then similar to that  of a square of a  bosonic annihilation operator.  
 
  {\bf iii)} An adjoint coherent phase where $H^\dagger$ has a continuous complex spectrum while $H$ has no bounded  eigenstates. 
  Here the normal modes are a pair of creation operators. While {\bf ii)} and {\bf iii)}  might be considered as having no proper 
  biorthogonal eigenstates, the convergent eigenstates (of $H$ or $H^\dagger$) constitute a generalization of the standard coherent 
  states, which  arise here in the particular case  of a non-diagonalizable matrix ${\cal MH}$.   
  These regimes may be considered to correspond to a broken generalized ${\cal PT}$ symmetry, since there are complex eigenvalues. 
  Nonetheless, the latter do not emerge from the coalescence of two or more real  eigenvalues  \cite{bender:2007:rpp} but from the 
  onset of convergence of eigenstates with complex quantum number $\nu$.  
  
We have also analyzed the transition curves between these previous regimes, where  one of the operators changes from creation to 
annihilation (or viceversa). At these curves 
such operator is actually a coordinate (or momentum), and even though there is just a discrete spectrum 
(with bounded eigenstates) of either $H$ or $H^\dagger$, the biorthogonality relations are still preserved. Explicit expressions for 
eigenfunctions were provided in all regimes.

 The normal mode decomposition  of the $N$-dimensional non-hermitian case has also been discussed, together with the corresponding 
 harmonic regime. It opens the way to investigate in detail along these lines 
 the spectrum of more complex specific non-hermitian quadratic systems. 
 
\appendix*
\section{Solutions of the Schr\"{o}dinger equation in the case of continuous spectrum}\label{sec:ap1}

The solutions to the Schr\"{o}dinger equation \eqref{eqn:schrodinger} can be obtained by making the substitution 
\begin{equation}
	\psi(x) = \exp\left[ -\frac{i\tilde{B}+\lambda}{2\tilde{A}_-}x^2  \right] \phi\left(\frac{x}{\gamma}\right).
	\label{eqn:susti}
\end{equation}
We obtain the Hermite equation \cite{lebedev:book}:
\begin{equation}
	\phi^{\prime\prime}(z) - 2 z \phi^\prime(z) + 2\nu \phi(z) = 0,
	\label{eqn:hermite}
\end{equation}
with $z = x/\gamma$ and $\nu = (2E-\lambda)/(2\lambda)$. 
For complex $\nu$, four solutions are:
\begin{equation}
	\begin{aligned}
		\phi_\nu^{(1)}(z) = H_\nu(z), &\quad \phi_\nu^{(2)}(z) = H_\nu(-z) \\
		\phi_\nu^{(3)}(z) = e^{z^2}H_{-\nu-1}(iz), &\quad \phi_\nu^{(4)}(z) = e^{z^2}H_{-\nu-1}(-iz),
	\end{aligned}
	\label{eqn:solus1}
\end{equation}
where $H_\nu$ are the Hermite functions \cite{lebedev:book}. 
Since the Hermite equation is of second order, any of these solutions can be written as a linear combination of two others. 
For instance, for real $A, B_+, B_- > 0$:
\begin{equation}
	\begin{aligned}
		H_\nu(z) &= \frac{2^\nu\Gamma(\nu+1)}{\sqrt{\pi}}e^{z^2} \left[ e^{\nu\pi i /2}H_{-\nu-1}(iz) \right .\\
		&\left . + e^{-\nu\pi i/2}H_{-\nu-1}(-iz) \right].
	\end{aligned}
	\label{eqn:relacion_hermite}
\end{equation}
Additionaly, note that for integer $\nu\geq0$, $\phi_1 = (-1)^\nu \phi_2$ whereas for integer $\nu<0$, $\phi_3 = (-1)^{\nu+1} \phi_4$.

The asymptotic behaviour of the Hermite functions for $|\arg z| < 3/4$ goes as follows:
\begin{equation}
	H_\nu(z) \sim (2z)^\nu + O(|z|^{\nu-2}), 
	\label{eqn:asinto1}
\end{equation}
and for $\pi/4 + \delta \leq \arg z \leq 5\pi/4 - \delta$ (which includes $z$ on the real negative axis):
\begin{equation}
	\begin{aligned}
		H_\nu(z) &\sim (2z)^\nu\left[ 1 + O(|z|^{-2}) \right]  -\\ 
		&\frac{\sqrt{\pi}e^{\nu\pi i}}{\Gamma(-\nu)}e^{z^2}z^{-\nu-1}\left[ 1 + O(|z|^{-2}) \right].
	\end{aligned}
	\label{eqn:asinto2}
\end{equation}
Note that:
\begin{equation}
	e^{z^2} \exp\left[ -\frac{i\tilde{B}+\lambda}{2\tilde{A}_-}x^2  \right]  = 
	\exp\left[ -\frac{i\tilde{B}-\lambda}{2\tilde{A}_-}x^2  \right].
\end{equation}

For hermitian $H$, $\tilde{B}$ is either a real number or zero, and $\lambda$ determines whether the eigenfunctions are bounded or 
not (i.e., if $\lambda$ is real and positive then there are \emph{some} bounded eigenfunctions, whereas for $\lambda$ negative or 
imaginary every eigenfunction is divergent).
In such case, for positive, integer $\nu$ only $\phi_\nu^{(1)}$ (and $\phi_{\nu}^{(2)}$, since they are linearly dependent) may be
 bounded (see Eq. \eqref{eqn:asinto2}), and for other values of $\nu$ there are no bounded eigenfunctions.
On the other hand, for non-Hermitian $H$, the convergence of both linearly independent eigenstates may be assured provided 
that $\Re[(i\tilde{B}-\lambda)/\tilde{A}_-] > 0$, which is fulfilled in region {\bf II}, i.e., when both $b$ and $\brd$ have convergent vacua.
Moreover, both linearly independent eigenstates may be convergent even if $\lambda$ is an imaginary number or zero, which implies 
for real $A, B_{\pm}$, that region {\bf II} extends into the imaginary part of the spectrum in Fig. \ref{fig1}.

The eigenfunctions of $H$ must then be constructed from \eqref{eqn:solus1} in such a way that they behave as the eigenstates $\Ket{n_b}$ 
and $\Ket{n_{\brd}}$, i.e., they satisfy Eqs. \eqref{sts} and \eqref{sts2}, and they must be even or odd with respect to coordinate 
inversion $x \rightarrow -x$ (since the Hamiltonian is parity invariant).
These considerations lead to the eigenfunctions \eqref{eqn:solu1gral} and \eqref{eqn:solu2gral}.

\acknowledgments
The authors acknowledge support from
CONICET (JG and Grant No. PIP 112201501-00732) and
CIC (R.R.) of Argentina.

\end{document}